\documentclass{PoS}

\title{Evolution of Gas in Galaxies}

\ShortTitle{Gas Evolution}

\author{\speaker{Lister Staveley-Smith}\thanks{West Australian 
Premier's Fellow}\\
        School of Physics, University of Western Australia, 
        Crawley, WA 6009, Australia\\
        E-mail: \email{Lister.Staveley-Smith@uwa.edu.au}}

\abstract{The SKA will be a unique instrument with which to study the
evolution of the gas content of galaxies. A proposed deep ($\sim$ 8
Msec) `pencil-beam' survey is simulated using recently updated
specifications for SKA sensitivity and survey speed. Almost $10^7$
galaxies could be detected in the redshifted 21cm line, most at
redshifts in excess of two.  This will enable confident statements to
be made about the evolution of the cosmic HI density and the HI mass
function to $z=3$, corresponding to a lookback time of 11
Gyr. However, galaxies or groups of galaxies with masses the same as
the most HI-massive galaxies at $z=0$ will be detectable at redshifts
of 6, if they exist. The ideal instrument for studying HI evolution
would have an instantaneous sensitivity at least a factor of two
higher than current specifications in the critical frequency range
200-500 MHz, or $A/T>2\times 10^4$ m$^2$ K$^{-1}$.  The capabilities
of the SKA will be highly complementary to ALMA which will be able to
study the evolution of the molecular gas component over the same
redshift range.}

\FullConference{From planets to dark energy: the modern radio universe\\
		 October 1-5 2007\\
		 University of Manchester, Manchester, UK}

\begin{document}

\section{Introduction}

Understanding the evolution of galaxies is one of the goals of modern
cosmology and one of the five key science goals of the SKA
\cite{cr04}. A key ingredient in galaxy evolution is the gas supplied
through various accretion, merger and feedback processes that occur
during the assembly of galaxies. This gas, which is mainly hydrogen,
passes through a neutral atomic phase and later condenses into massive
molecular clouds and stars. During the atomic phase, it can readily be
traced with the 21cm hyperfine spin-flip transition and the
Gunn-Peterson effect.

At high redshifts, $z>1.6$, neutral hydrogen is currently traced by 
ground-based observations of
Gunn-Peterson absorption lines against bright background QSOs. Such
observations demonstrate that the bulk of the neutral hydrogen is in the
Damped Ly-$\alpha$ systems with column densities exceeding $2\times10^{20}$ 
cm$^{-2}$. However, interpretation of these observations tends hampered by
insufficient lines of sight and serious uncertainties associated with
dust obscuration, gravitational lensing and intrinsic source size. These
uncertainties lead to contradictory results from measurements associated
with bright QSOs, faint QSOs and Gamma Ray Bursts \cite{Jas05}
\cite{Pro06} \cite{Por07}. However, with the SKA, galaxies will be detected
at similar redshifts in 21cm line emission, which will lead to a clearer
understanding of the distribution of gas in the Universe, and the manner
in which the gas content of galaxies evolves with time.

In order to measure the gas content of galaxies at the highest redshift,
a deep pencil beam survey with the SKA is proposed. Currently proposed SKA
specifications \cite{ska07} are adopted, and used to generate artificial
galaxy catalogues which are used to simulate the accuracy with which one
simple parameter can be recovered, namely the cosmic HI density -- 
the comoving volume density of neutral hydrogen.

\section{An SKA Deep Field}

An early goal in the design and development of the SKA has been the
requirement to detect and image galaxies beyond a redshift of unity in
the redshifted 21cm line of neutral hydrogen and in the radio
continuum \cite{vdH04}. However, the astronomy community has since
developed an impressive list of complementary science goals for the
SKA, including the study of the early intergalactic medium, dark
energy and cosmology, pulsars and tests of gravitational theory,
planet formation and cosmic magnetism \cite{cr04}. The resultant range
of requirements (e.g. frequencies, baselines, field-of-view) has
inevitably added to the cost of the SKA and created complexities in
developing a design. A recent SKA project study \cite{ska07} has
recommended the adoption of a preliminary set of specifications for
the SKA and its various phases. For the purpose of the discussion of a
putative SKA deep field, these specifications are adopted in this
paper. Where multiple specifications exist due to technology
uncertainty, the specifications relating to the simplest technology
has been chosen.  The SKA specifications, although realistic, are 
nevertheless still indicative and are
likely to change as the SKA design evolves. For reference, a
summary of the adopted values, in the frequency range of interest, is
listed in Table~\ref{t:lssspecs}.

\begin{table}[h]
\begin{center}
\begin{tabular}{lccccc}
\hline
Frequency & $z$ & $A/T^1$ & $\Omega$ & $\tau$ &  $N$ \\
MHz & & m$^2$ K$^{-1}$ & deg$^2$ & s \\ \hline
70 -- 200 & 6.1--19 & 3000 & 200 & $8\times10^6$ & 0 \\
200 -- 500 & 1.8--6.1 & 7500 & 200 & $8\times10^6$ & $6.6\times10^6$ \\
500 -- 1000$^2$ & 0.42--1.8 & 9000 & 2.0 & $8\times10^6$ & $4.4\times10^5$ \\
1000 -- 1420$^2$ & 0--0.42 & 9000 & 0.4 & $8\times10^6$ & $2.1\times10^4$ \\ \hline
\end{tabular} 
\end{center}
{\scriptsize
$^1$Assumes that 75\% of the SKA aperture will reside on baseline
lengths short enough not to significantly resolve distant galaxies.\\
$^2$Assumes wideband feed technology which gives a slightly better
sensitivity ($A/T$), but a dramatically worse field of view.
}

\caption{The adopted set of SKA specifications used in the
simulations, and the number of galaxies detectable in the simulations,
in a deep-field HI survey. $z$ is the redshift range; $A/T$ is the telescope
gain; $\Omega$ is the field of view; $\tau$ is the integration time;
$N$ is the number of galaxy detections expected. The specifications
follow the current SKA project office recommendations \cite{ska07},
but are subject to modification. There are a number of
other assumptions made in simulating the numbers of galaxies detected
in this survey, the most important ones being: that the comoving HI
mass function and velocity function is described by HIPASS \cite{z05};
that the galaxy detection threshold is 5-$\sigma$, and that the
minimum resolution used for the detection algorithm is 0.1 MHz.}
\label{t:lssspecs} 
\end{table}

A critical requirement for the study of galaxy evolution is the
ability to detect galaxies at the highest redshift. Since the
bandwidth is defined by the Doppler width of the galaxies, this
ability is only dependent on the instantaneous sensitivity $A/T$, and the
integration time $\tau$. Unlike for many other science goals, `survey'
or `mapping' speed is much less important as long as the field-of-view
is able to deliver suitably large statistical samples. High mapping
speed, on its own, is insufficient to ensure that the highest redshift
galaxies can be detected.

For this paper, the sole question of the evolution of the cosmic HI
density is addressed. This only requires the detection, not imaging,
of galaxies. Therefore, somewhere between 50\% (baselines less than 5
km) and 75\% (baselines less than 180 km) of the full SKA collecting
area is available, before galaxies become resolved and more difficult
to detect. Here, it is assumed that 75\% is available, corresponding
to a sensitivity of $A/T = 7500$ m$^2$ K$^{-1}$ in the critical
redshift range 1.8 -- 6.1. Values for other frequencies are listed in
Table~\ref{t:lssspecs}. Low redshift galaxies remain
heavily resolved. Nevertheless, at the higher redshifts of interest in
this paper, where the great majority of the galaxies will be detected,
the assumption remains reasonable. Available integration time is more
difficult to judge but, given the wide range of science possibilities
opened up with a deep field observation, it seems likely that a
substantial integration time may be feasible. It is assumed that 8
Msec, which corresponds to $180 \times 12$ hrs, will be available over
the first few years of operation of the SKA.

\begin{figure}[ht]
\begin{center}
\includegraphics[width=0.8\textwidth]{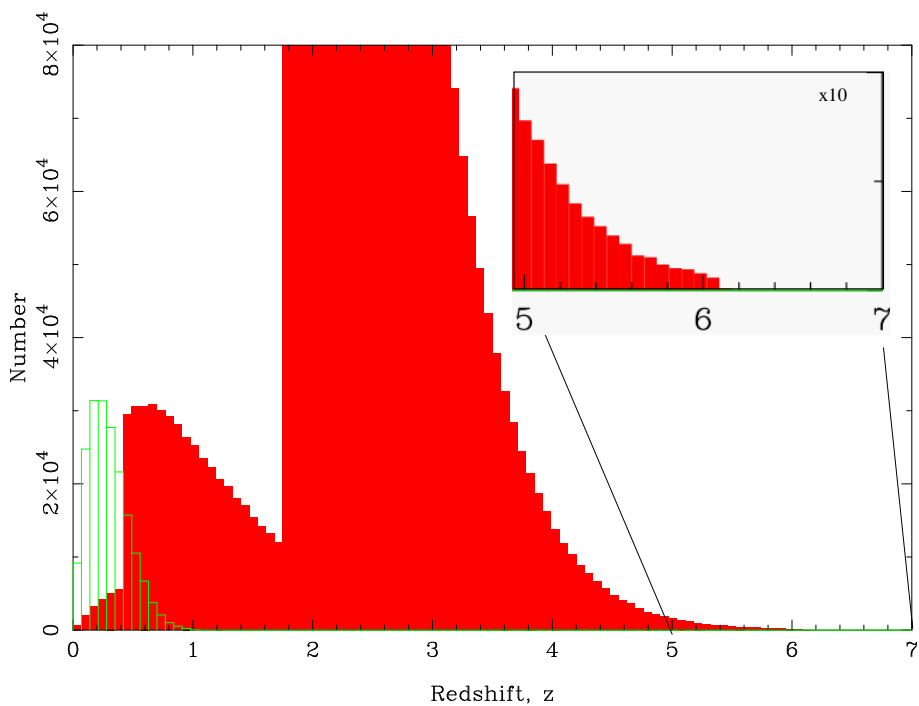}
\end{center}
\caption{Solid (red) histogram: the number of galaxies expected to be 
detected in HI in different redshift bins from 0 to 7 in a deep SKA 
`pencil beam' survey of integration time 8 Msec; open (green) histogram: the
number of galaxies expected below $z=1$ in the proposed ASKAP deep survey
(Johnston et al. 2007). Each redshift bin is of width 
$\Delta z=0.08$ and the total numbers of galaxies
in different redshift intervals is listed in the previous table. 
The histogram is truncated due to the large numbers of galaxies near $z=2$.}
\label{f:lsshisto}
\end{figure}

\begin{figure}[ht]
\begin{center}
\includegraphics[width=0.8\textwidth]{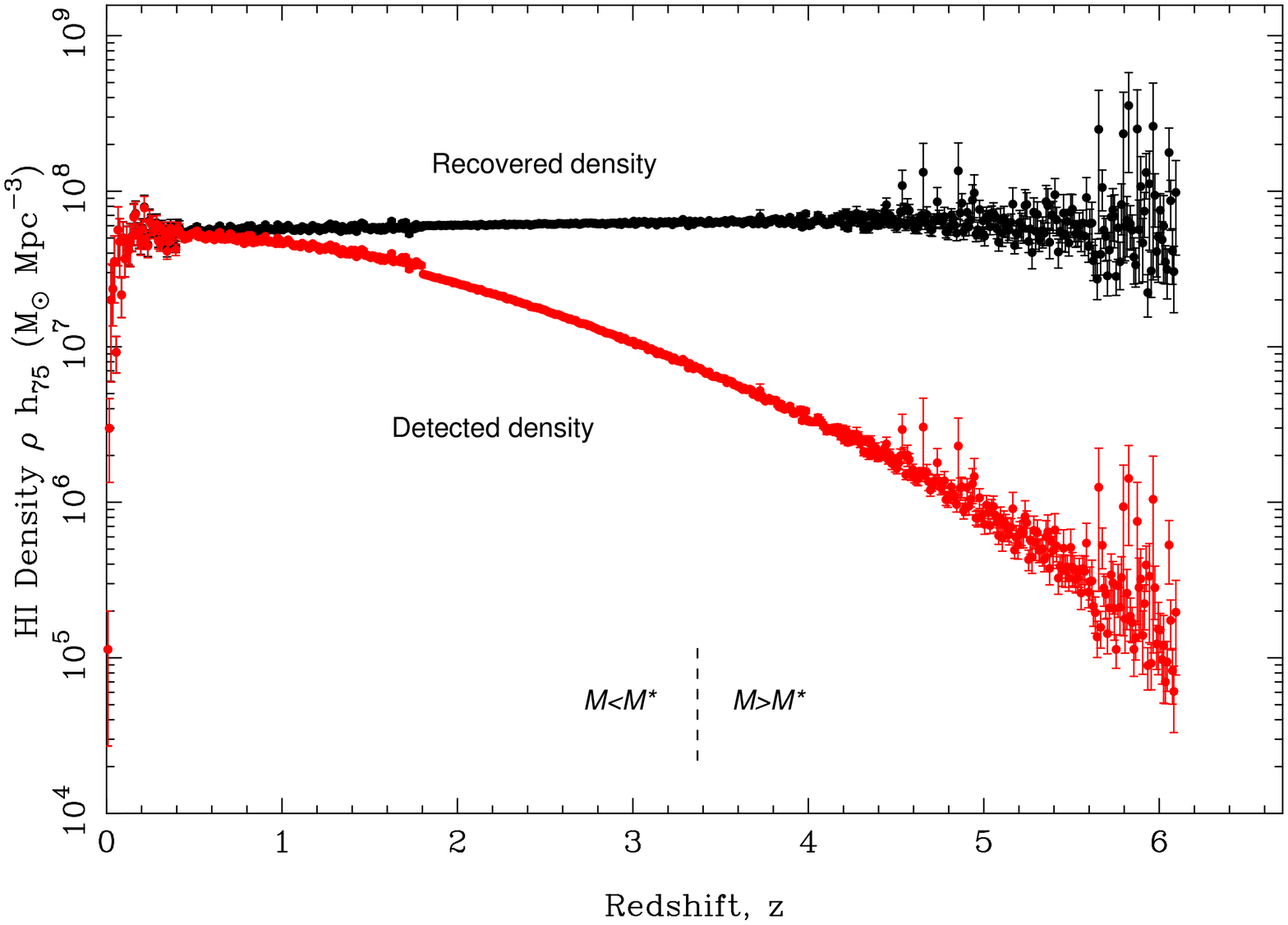}
\end{center}
\caption{A simulation which shows how well the cosmic HI density can be
recovered at various redshifts using the simulated galaxy catalogue of
the previous figure. The upper (black) points show the recovered density,
after correction for galaxies below the detection threshold; the lower (red)
points show the measured density of directly detected galaxies only.}
\label{f:lssdens}
\end{figure}

\section{Results}

A simulated catalogue was created using the above telescope
specifications and a non-evolving HIPASS HI mass function \cite{z05}.
The redshift distribution is shown in Figure~\ref{f:lsshisto}. The
small field-of-view at frequencies from 500 to 1420 MHz results in a
relatively small number of objects detected at low redshifts. For
example, only 21,000 galaxies are detected at redshifts beneath 0.42,
which is a factor of 10 less than expected from the proposed ASKAP
deep survey (\cite{Joh07} and Figure~\ref{f:lsshisto}), despite the
latter being conducted with a telescope of less than 1\% of the SKA
collecting area.  However, a shallow SKA survey over a large area of
sky is a better way of detecting larger numbers of low-redshift
galaxies.  At higher redshifts, much larger numbers of galaxies are
detected, mainly due to the much larger cosmic volume surveyed. $M^*$
galaxies are seen to redshifts of about 3.4, beyond which the number of
detected galaxies drops off dramatically.

For each redshift interval, the HI density integral $\int M_H \phi(M_H) dM_H$
is calculated for the detected galaxies and shown in Figure~\ref{f:lssdens}.
The `detected' density remains within a factor of a few of the input density
at redshift up to $\sim 2$, but drops to 10\% at a redshift of $\sim 3.5$ as
the bulk of the mass-bearing galaxies are too faint to detect. Nevertheless,
using an assumption that the shape of the mass function is constant, it is
possible to recover the input mass density even at redshift 5 
(Figure~\ref{f:lssdens}) before shot noise errors become large. In 
practice, the shape of the mass function will change with redshift in an
unknown manner, so it will be hard to reliably recover the cosmic HI mass
density much beyond a redshift of 3 from emission measurements alone.

\section{Discussion}

This simulation has demonstrated the large numbers of high-redshift
galaxies that can be detected by the SKA in a significant, but
feasible, HI survey of galaxies, and has demonstrated the high
accuracy with which the cosmic HI density can be measured. Whilst
galaxy numbers are low at redshifts below 1.8, this is largely due to
the small field of view available to single pixel technology. Adoption
of widefield detection technology has the potential to greatly
increase numbers. Although the subsequent reduction of errors
associated with shot noise and cosmic variance is important for many
science goals, greater gains in the field of galaxy evolution study
are likely to be made with better instantaneous sensitivity at lower
frequencies. This will increase the ability to detect changes in the
HI mass function and reduce the density extrapolation required to
account for low-mass galaxies at redshifts approaching 3. An
appropriate goal of $A/T>2\times 10^4$ m$^2$ K$^{-1}$ is therefore
suggested for frequencies below 500 MHz.

\end{document}